\documentclass[
    final            
  ,numberedheadings
  ]
  {aipproc}

\layoutstyle{8x11single}
\setcitestyle{numbers}
\def\beq{\begin{equation}}
\def\eeq{\end{equation}}
\def\bea{\begin{eqnarray}}
\def\eea{\end{eqnarray}}
\def\beqa{\begin{equation}\begin{array}{l}}
\def\eeqa{\end{array}\end{equation}}
\def\eqlab#1{\label{eq:#1}}
\def\figlab#1{\label{fig:#1}}

\def\Eqref#1{Eq.~(\ref{eq:#1})}

\def\Figref#1{Fig.~\ref{fig:#1}}

\def\al{\alpha}
\def\be{\beta}
 
\def\de{\delta} \def\De{\Delta}
  \def\eps{\epsilon}

\def\si{\sigma} 
\def\th{\theta}

\def\nn{\nonumber}

\def\im{\mbox{Im}}

\begin{document}

\title{Proton polarizabilities: status, relevance, prospects}

 

\author{Vladimir Pascalutsa}{
  address={
  Institut f\"ur Kernphysik, Johannes Gutenberg Universit\"at Mainz,  Germany}
}

\begin{abstract}
This is a brief review of the status of understanding
the proton polarizabilities in chiral perturbation theory
and of their relevance to the ``proton charge radius puzzle".
\end{abstract}

\maketitle


\section{Introduction}
A nucleon immersed in an external electromagnetic field acquires 
the electric and magnetic dipole moments which size is
given, respectively, by the electric and magnetic {\it polarizabilities}
$\al_{E1}$ and $\beta_{M1}$. These static quantities,
together with quantities such as the anomalous magnetic moment
and charge radius, reflect the complexity of the nucleon
structure. Their empirical determination is very important for at least two reasons:
first is that we would like to test our understanding of the nucleon structure
based on microscopic calculations of these quantities,
and second is that it enables us to evaluate the polarizability effects
in phenomena such as hydrogen Lamb shift, properties of nuclear matter, etc.
Here I shall illustrate these two motivations by looking at the description
of proton polarizabilities in chiral perturbation theory (Sect.~2), and at their relevance
to the muonic-hydrogen Lamb shift measurement (Sect.~3). I shall conclude with some words on prospects 
for better measurements of proton polarizabilities and of their momentum-transner
dependencies.   
  
\section{Status}
In Chiral Perturbation Theory (ChPT)~\citep{Weinberg:1978kz,Gasser:1983yg},
 the nucleon polarizabilities should largely come as a prediction
since the leading chiral-loop contribution is of order $p^3$, while
the unknown low-energy constants (LECs) come only  at order $p^4$. Recall 
that $p$ is of order $m_\pi /(4\pi f_\pi)$ and hence $p^4$ contribution is expected to be no greater than
15 percent of $p^3$. 
The result, however, depends on how or whether one includes the relativistic effects as
well as the effects due to the $\Delta(1232)$-isobar excitation. The leading relativistic corrections
carries an extra factor of $m_\pi/M_N$ and hence {\it nominally} is of order $p^4$. The scheme which consistently shuffles
the relativistic corrections to the order where they nominally should appear is called Heavy-Baryon ChPT (HBChPT)
\citep{JeM91a}, in
contradistinction with Baryon ChPT (BChPT) which simply follows from the manifestly Lorentz-invariant
Lagrangian of ChPT with baryons fields~\citep{GSS89,Fuchs:2003qc}.
The leading $\Delta(1232)$ contribution to polarizabilities
is of order $p^4/\Delta$, where $\De= M_\De-M_N \approx 300$ MeV is the Delta-nucleon mass difference
and hence, depending on counting, had been considered  to be of order $p^3$ ($\eps$-expansion \citep{Hemmert:1997ye}) or $p^4$
(``resonance saturation"), or in between ($\de$-expansion \citep{Pascalutsa:2003zk}).
By now, all the relevant contributions have been calculated in both HBChPT and BChPT and their
numerical values are given as follows (in units of $10^{-4}$ fm$^3$):
\bea
&& 0(p^3) \mbox{ BChPT~\citep{Bernard:1991rq}:}\quad \al_{E1} = 6.8,\quad  \be_{M1} = -1.8\,; \qquad 
O( p^3)
\mbox{ HBChPT~\citep{Bernard:1995dp}:}\quad \al_{E1} = 12.2,\quad  \be_{M1} = 1.2.\nn \\
&& O\Big(\frac{p^4}{\De}\Big) \mbox{ BChPT~\citep{Lensky:2009uv}:} \quad\al_{E1} = 4.0,  \quad\be_{M1} = 5.8\,;\qquad
O\Big(\frac{p^4}{\De}\Big)\mbox{ HBChPT~\citep{Hemmert:1996rw}:} \quad\al_{E1} = 8.6,  \quad \be_{M1} = 13.5\,; \nn\\
&& \mbox{$O(p^3+p^4/\Delta)$ BChPT}: \,\, \quad\al_{E1} = 10.8,  \quad\be_{M1} = 4.0\,;\qquad
\mbox{$O(p^3+p^4/\Delta)$ HBChPT:} \quad\al_{E1} = 20.8,  \quad \be_{M1} = 14.7\,
\nn
\eea
with
a relatively small uncertainty due to higher-order ($p^4$) contribution. Modern evaluations of the Baldin
sum rule \citep{Baldin}
yield for the sum of polarizabilities a value of 13.8(5) which compares well with either the total 
  $O(p^3+p^4/\Delta)$ BChPT value or with $O(p^3)$ HBChPT value. This shows that in HBChPT the 
  $\De$ contributions should only be treated together with $O(p^4)$. If 
   the deference between the BChPT and HBChPT numbers comes indeed from
  recoil corrections, then they are too significant to be neglected, and hence $O(p^4)$ is to be mandatorily
  included in HBChPT. In case of proton Compton scattering, where these polarizabilities prominently appear, 
  the calculations show that upon inclusion of $O(p^4)$ contributions the HBChPT achieves roughly the same 
  results as $O(p^3+p^4/\Delta)$ BChPT \citep{Lensky:2012ag}, albeit with a loss of some predictive power due to the appearance
  of two new LECs. 
  
The present status of the BChPT, HBChPT, as well as ``more empirical" extractions of proton polarizabilities,
as summarised  in \citep{Krupina:2013dya}, is shown in \Figref{potato}.
\begin{figure}[t]
\includegraphics[width=0.5\linewidth]{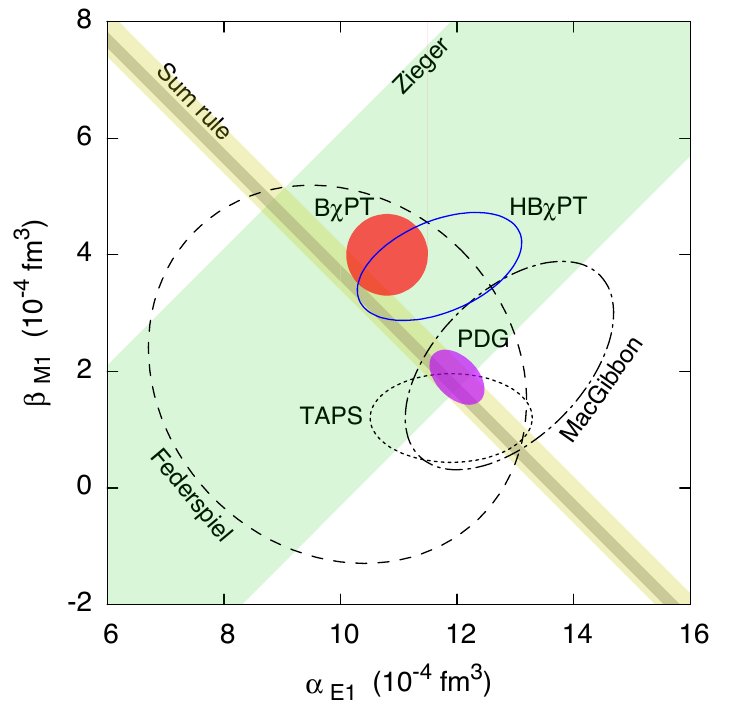}
\caption{(Color online). The scalar polarizabilities of the proton. 
Magenta blob represents the PDG summary~\citep{Beringer:1900zz}.
Experimental
results are from Federspiel et~al.~\citep{Federspiel:1991yd},
Zieger et al.~\citep{Zieger:1992jq}, MacGibbon et al.~\citep{MacG95},
and TAPS~\citep{MAMI01}.
`Sum Rule' indicates the Baldin sum rule evaluations of 
$\alpha_{E1}+\beta_{M1}$~\citep{MAMI01} (broader band) and \citep{Bab98}.
ChPT calculations are from \citep{Lensky:2009uv} (BChPT---red blob)
and the `unconstrained fit' of \citep{McGovern:2012ew} (HBChPT---blue ellipse).} 
\figlab{potato}
\end{figure}
Note the significant discrepancy of the BChPT prediction with the current Particle Data Group values, which thes far
has been attributed to a sizeable underestimate of uncertainty in the TAPS and subsequently PDG values. 

\section{Relevance: hydrogen Lamb shift}
The electric polarizability of the proton is responsible for a zero-range force in atoms, which lead to a shift in the $S$-levels:
\bea
\eqlab{LS}
\De E_{nS}^{(\mathrm{pol.})} = -4 \al_{em} \, \phi_n^2(0) \int\limits_0^\infty d Q \left[\sqrt{1+\frac{Q^2}{4m_\ell^2} } -\frac{Q}{2m_\ell}\right]
\al_{E1} (Q^2),
\eea
where $\al_{em}$ is the fine-structure constant, 
$\phi_n^2(0) = \al_{em}^3 m_r^3/(\pi n^3) $ is the square of the hydrogen wave-function at the origin, $m_\ell$
is the lepton mass and $m_r$ is 
the reduced mass: $m_r = M_p m_\ell/(M_p+m_\ell)$. 
The effect of magnetic polarizability is suppressed.

The effect in \Eqref{LS} is of order $\al_{em}^5$; there is one $\al_{em}$ implicit in the polarizability. It is therefore
of the same order as the effects of 3rd Zemach radius and can make an impact on "charge radius puzzle" 
\citep{Pohl:2010zza,Antognini:1900ns}, i.e.,
the 7$\si$ discrepancy between the proton charge radius extraction based on either the electronic ($e$H) or muonic ($\mu$H)
hydrogen Lamb shift. 
The factor in the square brackets of 
\Eqref{LS} acts a soft cutoff at the scale of order of the lepton mass $m_\ell$, and hence the proton polarizability
contribution in $\mu$H is expected to be bigger than in $e$H. How much bigger?

The transfer-momentum
dependence of $\al_{E1}$ is inferred from the forward doubly-virtual Compton scattering, and hence is not
accessible in a direct experiment. Only the sum, $\al_{E1}(Q^2) + \be_{M1}(Q^2)$, is accessible through
a generalized Baldin sum rule. The Baldin sum rule has been evaluated in several works leading to 
the so-called `inelastic' contribution to the shift $\mu$H 2S level~\citep{Pachucki:1999zza,Faustov:2001pn,Carlson:2011zd,Birse:2012eb}:
\beq
\De E_{2S}^{(\mathrm{inel.})} =
-4 \al_{em} \, \phi_2^2(0) \int\limits_0^\infty d Q \left[\sqrt{1+\frac{Q^2}{4m_\mu^2} } -\frac{Q}{2m_\mu}\right]
\left\{ \al_{E1} (Q^2)+  \be_{M1} (Q^2)\right\} 
\approx   - 13 \, \, \mu\mbox{eV}.
\eeq
One then need subtract $\be_{M1}(Q^2)$ to obtain the energy shift as defined in \Eqref{LS},
i.e.: $\De E_{2S}^{(\mathrm{pol.})} = \De E_{2S}^{(\mathrm{inel.})} + \De E_{2S}^{(\mathrm{subt.})}$, with
 \beq
\De E_{2S}^{(\mathrm{subt.})} =
4 \al_{em} \, \phi_2^2(0) \int\limits_0^\infty d Q \left[\sqrt{1+\frac{Q^2}{4m_\mu^2} } -\frac{Q}{2m_\mu}\right]
  \, \be_{M1} (Q^2)\,.
  \eqlab{Subt}
\eeq
In other words the problem is shifted to finding $\be_{M1}(Q^2)$ which seems to be
just as unknown as $\al_{E1}(Q^2)$. This uncertainty of the polarizability contribution has been
exploited by Miller~\citep{Miller:2012ne} to suggest that it could be as large as  $-310$  $\mu$eV needed to resolve
the charge radius puzzle.

An insight can be gained by using ChPT, which should work well for momenta probed in atomic systems.
Based on general analytic properties of the momentum-transfer  dependence (i.e., analyticity in
the complex $Q^2$ plane, except for the negative real axis---timelike region) infer a
dispersion relation of the type:
\beq
\eqlab{theDR}
\left\{\begin{array}{c}\al_{E1} (Q^2)\\
\be_{M1}(Q^2)
\end{array}\right\} = \frac{1}{\pi}\int\limits_0^\infty   \frac{dt}{t+Q^2-i0^+} \, \im\, \left\{\begin{array}{c}\al_{E1}(-t) \\
\be_{M1}(-t)
\end{array}\right\},
\eeq
where $0^+$ is an infinitesmal positive number. An explicit $p^3$ calculation in HBChPT yields:
\beq
\im\,\be_{M1}^{(3)}(-t) =
\frac{\al_{em} g_A^2}{16 f_\pi^2} \, \frac{ m_\pi^2}{t^{3/2}}\,
\th(t-4m_\pi^2),
\eeq
where $g_A\simeq 1.27$, $f_\pi\simeq 92.4$ MeV are respectively the axial
and pion-decay constant; $m_\pi$ is charged-pion mass.
Substituting this into \Eqref{theDR} we obtain
\beq
\be_{M1}^{(3)}(Q^2) =
\frac{\al_{em} g_A^2}{16\pi f_\pi^2} 
\frac{m_\pi}{Q^2} \left[1- \frac{2 m_\pi}{Q} \arctan \frac{Q}{2m_\pi} \right],
\eeq
which reproduces the result of Birse and McGovern~\citep{Birse:2012eb}.
Substituting this into \Eqref{Subt}
and setting for simplicity $m_\mu = m_\pi$, we obtain the following subtraction contribution:
\beq
\De E_{2S}^{(\mathrm{subt.})} =  \frac{\al_{em}^5 m_r^3  g_A^2}{2(4\pi f_\pi)^2} 
\Big( \mbox{$\frac18$}- \mbox{$\frac14$} C + \mbox{$\frac13$} \ln 2\Big) \simeq 1.4 \,\, \mu\mbox{eV},
\eeq
where $C\simeq 0.9160$ is the Catalan number.
We conclude that the outcome is tiny and is very unlikely to change by orders of magnitude upon
refining the ChPT calculation. A calculation in BChPT is nevertheless forthcoming \citep{Alarcon}.

\section{Prospects}

There is a still big room for improvement of our knowledge of nucleon polarizabilities, 
most notably in the empirical knowledge 
of their static values, as well as of their momentum-transfer dependence. Even
the static electric and magnetic polarizabilities of the proton are not pinned down
to the accuracy of less than 10 percent. Such accuracy seems well within the reach
of current experimental capabilities and we shall definitely see a progress in this direction
in a very near future. The beam asymmetry of Compton scattering looks especially promising 
for a precise determination of the small magnetic polarizability~\citep{Krupina:2013dya}.
A new round of real- and virtual-Compton scattering experiments at low energies
on the proton and light nuclei targets has recently commenced at Mainz Mictrotron (MAMI)
at the University of Mainz. There is real-Compton scattering program 
running at the High Intensity Gamma Source (HIGS) facility at Duke
University. The new high-intensity beam facilities, such as MESA, will bring new opportunities in
this field. 

\section*{Acknowledgements}
The work is partially supported by the Deutsche Forschungsgemeinschaft (DFG) through Collaborative 
Research Center SFB 1044 ``The Low-Energy Frontier of the Standard Model".


\end{document}